\documentclass[%
10pt,
 amsmath,amssymb,
 longbibliography,
 aps,
 prd, 
]{revtex4-2}

\usepackage{graphicx}
\usepackage{bm}
\usepackage{hyperref}
\usepackage{subfigure}

\usepackage[nameinlink,noabbrev]{cleveref}
\usepackage[T1]{fontenc}
\crefname{equation}{Eq.}{Eqs.}
\crefname{section}{Section}{Sections}
\crefname{figure}{Figure}{Figures}
\crefname{table}{Table}{Tables}
\crefname{appendix}{Appendix}{Appendices}
\Crefname{figure}{Figure}{Figures}
\Crefname{equation}{Equation}{Equations}
\Crefname{section}{Section}{Sections}
\Crefname{table}{Table}{Tables}

\newcommand{\jcap}{JCAP}
\newcommand{\aap}{A\&A}
\newcommand{\mnras}{MNRAS}
\newcommand{\apjl}{ApJL}

\newcommand{\lcdm}{$\Lambda$CDM} 
\newcommand{\wowa}{$w_0w_a$CDM} 
\newcommand{\rd}{r_\mathrm{d}}
\newcommand{\Mpc}{\, \text{Mpc}}
\newcommand{\kmsMpc}{\,{\rm km\,s^{-1}\,Mpc^{-1}}}

\newcommand{\ob}{\omega_\mathrm{b}}
\newcommand{\ocdm}{\omega_\mathrm{cdm}}
\newcommand{\dchisq}{\Delta\chi^2_\mathrm{MAP}}

\begin{document}

\preprint{APS/123-QED}

\title{Persistent and serious challenge to the $\Lambda$CDM throne: Evidence for dynamical dark energy rising from combinations of different types of datasets}
\author{Mustapha Ishak$^{1}$}
\email{mishak@utdallas.edu}
\author{Leonel Medina-Varela$^{1,2}$}

\affiliation{
$^1$ Department of Physics, The University of Texas at Dallas, Richardson, Texas 75080, USA\\
$^2$ McWilliams Center for Cosmology and Astrophysics, Department of Physics, Carnegie Mellon University, Pittsburgh, PA 15213, USA \\
}

\begin{abstract}
We derive multiple constraints on dark energy and compare dynamical dark energy models with a time-varying equation of state ($w_0 w_a$CDM) versus a cosmological constant model ($\Lambda$CDM). We use Baryon Acoustic Oscillation (BAO) from DESI and DES, Cosmic Microwave Background from Planck with and without lensing from both Planck and ACT (noted CMBL and CMB, respectively), supernova (SN), and cross-correlations between galaxy positions and galaxy lensing (i.e. 3x2pts) from DES. 
First, we use pairs or trios of datasets where we exclude one type of dataset each time and categorize them as ``NO SN", ``NO CMB" and ``NO BAO" combinations. In all cases, we find that the data combinations favor the $w_0 w_a$CDM model over $\Lambda$CDM, with significance ranging from 2.3$\sigma$ to 3.3$\sigma$. For example, DESI+DESY6BAO+CMB yields 3.2$\sigma$ without SN, DESI+DESY6BAO+DESY5SN yields 3.3$\sigma$ without CMB, and CMB+DESY5SN+DES3x2pts yields 2.6$\sigma$ without BAO. 
The persistence of this pattern across various dataset combinations even when any of the datasets is excluded (along with their possible systematics) supports an overall validation of this trending result regardless of any specific dataset. 
Next, we use larger combinations of these datasets after verifying their mutual consistency within the $w_0 w_a$CDM model. We find combinations that give significance levels $\sim$4$\sigma$, with DESI+DESY6BAO+CMBL+DESY5SN reaching 4.4$\sigma$. 
In sum, while we need to remain prudent, the combination of the first step that supports a validation of the pattern of these results beyond any single type of dataset and their associated systematics, together with the second step showing high-significance results when such datasets are combined, presents a compelling overall portrait in favor of a dynamical dark energy with a time-evolving equation of state over a cosmological constant, and constitutes a serious challenge to the $\Lambda$CDM model's reign.

\end{abstract}

\maketitle

\vspace{-16pt} 
\section{\label{sec:introduction}Introduction}
The discovery of the expansion of the universe nearly a century ago was a major contribution to building our current standard model of cosmology. Such an expansion is not a mystery and came as a consequence of the theory of gravity of Einstein, see e.g. \cite{weinberg2008cosmology,peebles2020cosmologys}. However, the acceleration of such an expansion came as a surprise in 1998 \cite{SupernovaCosmologyProject:1998vns,SupernovaSearchTeam:1998fmf} even if the notion of a cosmological constant term in Einstein's equations of gravity and its effect on spacetime evolution were well-developed and studied from a theoretical point of view. A plethora of observations and cosmological analyses in the following 27 years continued to deepen the work on cosmic acceleration and confirmed it beyond any doubts. It has also been established that such an acceleration can be caused by a dynamical component or energy in the universe whose specific nature we do not currently understand and was dubbed as dark energy.

This led to an even more puzzling picture of our standard model of the universe where around 5 percent of the matter-energy content is made of “normal” baryonic matter that we, and other celestial objects, are made from, around 25 percent is made of dark matter, and an overwhelming 70 percent of the content is made of the aforementioned dark energy.   

The big question thus becomes what is dark energy? Is it the well-known cosmological constant that goes into the equations of Einstein General Relativity, see e.g. \cite{Carroll:2000fy,Peebles:2002gy,Copeland:2006wr,Ishak:2005xp}? Is it another scalar-field based dark energy component, e.g.  \cite{Carroll:2000fy,Peebles:2002gy,Copeland:2006wr,Ishak:2005xp}? Or is cosmic acceleration due to some modification to General Relativity at cosmic scales that could be cast like an effective dark energy, e.g. \cite{Carroll:2000fy,Peebles:2002gy,Copeland:2006wr,Ishak:2005xp}?  To discriminate between these dark energy candidates, astrophysicists have been using observations to measure to a high-level of accuracy and precision the so-called equation of state (EOS) of dark energy parameter, noted as ``w''.  This is a variable that represents the ratio of the pressure of dark energy over its energy density and allows for more flexible models for cosmic acceleration than just the cosmological constant. This EOS may also be a function of time and can be modeled with a parameter $w_0$ representing its value at the present time and a second parameter $w_a$ representing minus the slope of its time evolution ~\cite{Chevallier:2001,Linder2003}.  For a cosmological constant, $w$ is simply a constant in time given by the number of minus one (i.e. $w_0=-1$ and $w_a=0$), but for other possible candidates of dark energy or for departures from general relativity these parameters can take other values and potentially include time variation. It is worth clarifying that even a constant EOS that is different from minus one does produce a dynamical dark energy density  --  a nuance sometime overlooked in the general literature so it is good to note it here. Therefore, measuring the equation of state of dark energy using observations has become the state of the art approach to discriminate between dark energy candidates. A lot of progress has been made on methods to constrain the EOS and a large number of surveys and experiments have been designed to constrain it, see e.g. \cite{DESI2016a.Science,DES:2020ahh,2015arXiv150303757S,Euclid:2024yrr,2009arXiv0912.0201L}.

However, for over 25 years after the discovery of cosmic acceleration, the data and results have been in general centered around the minus-one value of the cosmological constant and also have been lacking power to constrain meaningfully a time-varying equation of state. While there was some small-significance results here and there of a possible departure from a cosmological constant, see e.g.  \cite{Zhao:2017cud}, a first strong result came in 2024 from the Dark Energy Spectroscopic Instrument (DESI) giving tantalizing hints for a dynamical dark energy with a time-evolving EOS using its year-1 data referred to as Data Release 1 (DR1) when combined with data from the Cosmic Microwave Background (CMB) radiation and its lensing and supernova (SN) data, see e.g.  \cite{DESI:2024mwx, DESI:2025zpo, ishak2024desi}.  The evidence for such results grew even stronger using DESI year-3 data (referred to as DR2), see e.g. \cite{DESI:2025zgx,DESI:2025fii,DESI:2025gwf}.

Big results require big scrutiny. A large number of papers reacted to the DESI results by either confirming them and extending them to other models, or by questioning whether the new results of a preference for dynamical dark energy with a time-evolving equation of state over a cosmological constant are due to some systematic effects or caveats in the analysis framework or within a given dataset or the other, see for a partial list the references \cite{Tada:2024xau, Wang:2024qan, Yin:2024xau, Luongo:2024xau, Cortes:2024xau, Colgain:2024xqj, Wang:2024rjd, Berghaus:2024kra, Wang:2024xau, Wang:2024pui, Shlivko:2024llw, Dinda:2024kjf, Bhattacharya:2024xau, Ramadan:2024xau, Roy:2024kni, Gialamas:2024lyw, Notari:2024xau, Liu:2024gfy, Orchard:2024xau, Hernandez-Almada:2024xau, Pourojaghi:2024xau, Giare:2024gpk, Jiang:2024xau, Efstathiou:2024xau, Reboucas:2024xau, RoyChoudhury:2024xau, RoyChoudhury:2024wri, Dhawan:2024xau, Linder:2024rdj, Park:2024xau2, Notari:2024xau2, Gao:2024ily, Gao:2024xau, Fikri:2024klc, Fikri:2024xau, Tiwari:2024gzo, Tang:2024lmo, Zheng:2024xau, Odintsov:2024xau, Colgain:2024mtg, Lewis:2024xau, Koussour:2024xau, Sakr:2025fay, Yang:2025xau, Huang:2025xau, Wolf:2025xau, Giare:2025pzu, Sousa-Neto:2025gpj, Ormondroyd:2025exu, Khoury:2025txd, Ormondroyd:2025iaf, Brandenberger:2025hof, Nesseris:2025lke, Kessler:2025kju, You:2025uon, Shlivko:2025fgv,Cai:2025mas,Popovic:2025glk,Li:2025ops,Kou:2025yfr,Santos:2025wiv, DESI:2025wyn, Akarsu:2025dmj, Dinda:2025iaq, Wang:2025bkk, Mirpoorian:2025rfp, RoyChoudhury:2025dhe, Scherer:2025esj, Liu:2025mub, Chen:2025mlf, Chen:2025wwn, Efstathiou:2025tie, vanderWesthuizen:2025iam, Sabogal:2025jbo, Linder:2025zxb, Odintsov:2025jfq, Araya:2025rqz, Li:2025eqh, Herold:2025hkb, Lee:2025kbn, Chen:2025jnr, Hogas:2025ahb, Qiang:2025cxp, Lee:2025pzo, Chudaykin:2025aux, Silva:2025twg, Camarena:2025upt, Chaudhary:2025pcc, Wang:2025vtw, Yao:2025wlx, Arora:2025msq, Paul:2025wix, Toomey:2025xyo, Fazzari:2025lzd, Dinda:2025hiu, Park:2025fbl, RoyChoudhury:2025iis, Alam:2025epg, Artola:2025zzb, Reeves:2025xau, Yadav:2025vgo, Rezaei:2025vhb, Xu:2025nsn}. 
 
In this paper, we first proceed by combining datasets while excluding one type of data at a time, showing that the result of dynamical dark energy stands despite such exclusions, providing a validation of the trend, and clarifying that the result is persistent regardless of a specific single type of dataset or its related systematic effects. Next, we combine all the datasets we considered in this work and find new high-significance results in favor of a time-evolving dynamical dark energy over the \lcdm\ model, and then discuss the implications of the results from the two steps and their joint consideration.

The paper is organized as follows. 
\cref{sec:Formalism} provides the formalism and the datasets used in the analysis. 
\cref{sec:exclusion} presents results from pairs or trios of datasets while excluding one type of dataset each time. 
In \cref{sec:combining}, results are presented for all three dataset types combined. 
A conclusion is provided in \cref{sec:conclusion}.

\section{\label{sec:Formalism}methodology and data}

\subsection{\label{sec:backround}Background and formalism}
We assume a Friedmann-Lemaitre-Robertson-Walker (FLRW) cosmology where the content of the universe is made of baryons, cold dark matter and a dark energy component. For simplicity and also based on previous studies that found that the dark energy equation of state constraints were mainly unchanged when including spatial curvature, see e.g. \cite{DESI:2024mwx, DESI:2025zpo, DESI:2025zgx}, we use a spatially flat geometry. 

The constraints on dark energy in this work come primarily from the expansion history and distance measurements so we focus the formalism's description here on the homogeneous evolution with the Hubble function given by:
\begin{equation}
H(z) = H_0 \Big[\Omega_\mathrm{m}(1+z)^3 + \Omega_\mathrm{DE} \frac{\rho_{\mathrm{DE}}(z)}{\rho_{\mathrm{DE},0}} \Big]^{1/2}~.
    \label{eqn:hubble}
\end{equation}
where $\Omega_{\rm m}=\Omega_{\rm b}+\Omega_{\rm cdm}$ is the total matter energy density parameter including baryons plus cold dark matter, $\Omega_{\rm DE}$ is the dark energy density parameter, $\rho_{\mathrm{DE}}(z)$ is the dark energy density with $\rho_{\mathrm{DE},0}$ being its value at the present time (i.e. $z=0$).  Dark energy is also commonly parametrized by its equation of state parameter defined as the ratio of its pressure over its energy density, i.e. $w(z) \equiv P(z)/(c^2 {\rho}_\mathrm{DE}(z))$.  We employ the commonly used CPL parametrization of a time-evolving equation of state $w(z)$ that is given by ~\cite{Chevallier:2001,Linder2003}: 
\begin{equation}
    w(z)=w_0+w_a\,\frac{z}{1+z}= w_0 + w_a (1-a)~\,
    \label{eq:w0wa}
\end{equation} 
where $a = (1+z)^{-1}$ is the normalized scale factor. We refer to models with a time-evolving equation of state of dark energy using \cref{eq:w0wa} as \wowa\ and models with a cosmological constant, $\Lambda$, as \lcdm\ (i.e. with $w_0=-1$ and $w_a=0$).
We note that multiple analyses have shown that the result of preference of dynamical dark energy stands equally for other functional form parameterizations, binning methods, or Gaussian processes, see e.g. \cite{DESI:2025fii,DESI:2024kob,DESI:2024aqx}. The cosmological parameters used for the models considered and the corresponding priors used are described in \cref{tab:priors} in the appendix. 

For our cosmological analysis, we employ the cosmological inference software \texttt{cobaya} \cite{Torrado:2019, Torrado:2021}, incorporating likelihoods from DESI \cite{DESI:2025zgx}, Planck \cite{Planck-2018-likelihoods}-Camspec \cite{Rosenberg:2022}, ACT \cite{Madhavacheril:ACT-DR6, Qu:2023}, DES-SN5YR \cite{DES:2024jxu}. 
For likelihoods, we utilize public packages that are either part of the \texttt{cobaya} distribution or available from the respective research teams \footnote{we acknowledge and thank Lanyang Yi for sharing a likelihood for DESY3 for Cobaya}. We perform Bayesian inference using the Metropolis-Hastings MCMC sampler \cite{LewisMCMC:2002, LewisMCMC:2013} within \texttt{cobaya}, requiring a convergence criterion of $R-1 \leq 0.01$ for MCMC chains. 

We use \texttt{iminuit} \cite{iminuit} code as invoked from  \texttt{cobaya} and starting from the maximum a posteriori (MAP) points of each of the chains obtained from the MCMC sampling in order to determine the best fit points and the corresponding $\chi^2$ for a model. 
To compare the \wowa\ model fitting to the data versus the \lcdm\ model, we use the $\Delta\chi^2_\mathrm{MAP}\equiv-2\Delta\ln \mathcal{L}$ where $\ln \mathcal{L}$ is the log posteriors at the maximum posterior points. Since the \lcdm\ (null hypothesis model) is nested into the \wowa\ (alternate model with two extra parameters), Wilks' theorem \cite{Wilks:1938dza} states that the defined  $\Delta\chi^2_\mathrm{MAP}$ should follows a $\chi^2$ distribution with two degrees of freedom. We then translate this $\Delta\chi^2_\mathrm{MAP}$ into the commonly-used $n\sigma$ significance level of a one-dimensional Gaussian distribution (right-hand side of \cref{eq:CDF} below) equated to the cumulative distribution of our $\chi^2$ (left-hand side of \cref{eq:CDF}) as 
\begin{equation}
    \mathrm{CDF}_{\chi^2}\left(|\Delta\chi^2_\mathrm{MAP}|\, ;\, 2\,\mathrm{dof}\right) = \frac{1}{\sqrt{2\pi}}\int_{-n}^{n} e^{-t^2/2} dt~.
\label{eq:CDF}
\end{equation} 
The $n\sigma$ is then calculated from the inverse of \cref{eq:CDF}. Additionally,  we use for model comparison, the Deviance Information Criterion (DIC) difference, $\Delta(\mathrm{DIC}) = \mathrm{DIC}_{w_0w_a\mathrm{CDM}} -\mathrm{DIC}_{\Lambda\mathrm{CDM}}$ that penalizes the inclusion of extra parameters (i.e. $(w_0, w_a)$ here) and accounts for the Bayesian complexity of the models, see e.g. 
\cite{kass1995bayes,Liddle:2007fy,Trotta:2008qt,Grandis:2016fwl}.  

\subsection{\label{sec:datasets}Datasets}
\begin{itemize}
    \item {Baryon Acoustic Oscillations (BAO):} we use BAO distance data from DESI DR2 as described in \cite{DESI:2025zpo,DESI:2025zgx}. For the BGS tracer, we analyze the compressed $D_V/r_d$ measurements to obtain low redshift information within $0.1<z<0.4$. For other DESI tracers, we utilize the usual $D_M/r_d$ and $D_H/r_d$ measurements. Specifically, we consider two LRG (Luminous Red Galaxies) bins in the ranges $0.4<z<0.6$ and $0.6<z<0.8$, a combined LRG+ELG (Emission Line Galaxies) measurement within $0.8<z<1.1$, an ELG tracer measurement spanning $1.1<z<1.6$, and a QSO measurement within $0.8<z<2.1$. Systematic tests for BAO measurements from galaxy and quasar clustering are discussed in \cite{DESI:2025qqy}. Additionally, we include Ly$\alpha$ measurements within $1.8<z<4.2$, providing the highest redshift data point, see for description and validation \cite{DESI:2025zpo,DESI:2025qqu,DESI:2025brt}. This entire DESI dataset, covering redshifts from $0.1$ to \textit{$4.2$} and divided into seven main samples is referred to as BAO in this paper. We additionally include the BAO measurement from DESY6 \cite{DESY6BAO,DES:2025bxy} which utilizes a BAO optimized sample within $0.6<z<1.2$ with a $z_{eff}=0.851$. This data point is referred to as DESY6BAO in this paper. We note that there is a small overlap in the area and redshift range between the DES and DESI BAO but the correlation resulting from it is expected to be small, see e.g. \cite{DES:2025bxy}, and its exploration is beyond the scope of this paper. 

    \item {Type Ia Supernovae (SNeIa):} We use one of three SNeIa datasets at a time: PantheonPlus, Union3, or DESY5. The PantheonPlus dataset \cite{Brout:2022} includes 1550 spectroscopically-confirmed SNeIa within the redshift range $0.001<z<2.26$. The Union3 compilation \cite{Rubin:2023} consists of 2087 SNeIa spanning $0.01<z<2.26$, with 1363 overlapping with PantheonPlus, though the analysis techniques differ significantly. Lastly, the DESY5 dataset \cite{DES:2024tys} features 1635 photometrically-classified SNeIa with redshifts from $0.1<z<1.13$, supplemented by 194 low-redshift SNeIa, also included in PantheonPlus, within the redshift range $0.025<z<0.1$.

\item {Cosmic Microwave Background (CMB):} We incorporate temperature and polarization data from Planck \cite{Planck-2018-overview}. Specifically, we use the high-$\ell$ TTTEEE likelihood (\texttt{planck\_NPIPE\_highl\_CamSpec.TTTEEE}), along with low-$\ell$ TT (\texttt{planck\_2018\_lowl.TT}) and low-$\ell$ EE (\texttt{planck\_2018\_lowl.EE}) \cite{Planck:2019nip, Efstathiou:2021}, as implemented in \texttt{Cobaya}~\cite{Torrado:2021}. Additionally, we combine temperature and polarization anisotropies with CMB lensing data from the combination of NPIPE PR4 from Planck \cite{Carron:2022eyg,Rosenberg:2022} and the Atacama Cosmology Telescope (ACT) DR6 \citep{Madhavacheril:ACT-DR6,ACT:2023oei}. 

\item {Dark Energy Survey (3$\times$2-pt) (DESY3)}: We utilize data from the DES Year 3, which includes combined measurements of cosmic shear, galaxy-galaxy lensing, and galaxy clustering, referred to as the (3$\times$2-pt) analysis. Accordingly, we do not apply the Limber approximation for galaxy clustering on large angular scales but instead follow the precise method described in \cite{Fang:2019xat}. The DESY3 (3$\times$2-pt) analysis was performed using source galaxies in four redshift bins [0, 0.36, 0.63, 0.87, 2.0] and lens galaxies from the Maglim sample in the first four redshift bins [0.20, 0.40, 0.55, 0.70, 0.85, 0.95, 1.05]. We employed similar configurations for the \wowa\ models as in the DES paper \cite{DES:2022ccp} and refer the reader to this reference for further details.
   
\end{itemize}

\clearpage  

\vspace{-10pt} 

\section{\label{sec:exclusion}Validation by exclusion}
\vspace{-5pt} 

In a first step and before we show results from larger combinations of datasets further below, we show here constraints on the dark energy equation of state parameters from combining datasets in pairs of types where we exclude systematically one dataset type at a time and thus its possible associated systematics.

\vspace{-5pt} 

\subsection{\label{sec:no_SN}NO SN Data}
\vspace{-5pt} 

\begin{figure} 
    \centering
    \includegraphics[width = 0.48\columnwidth]{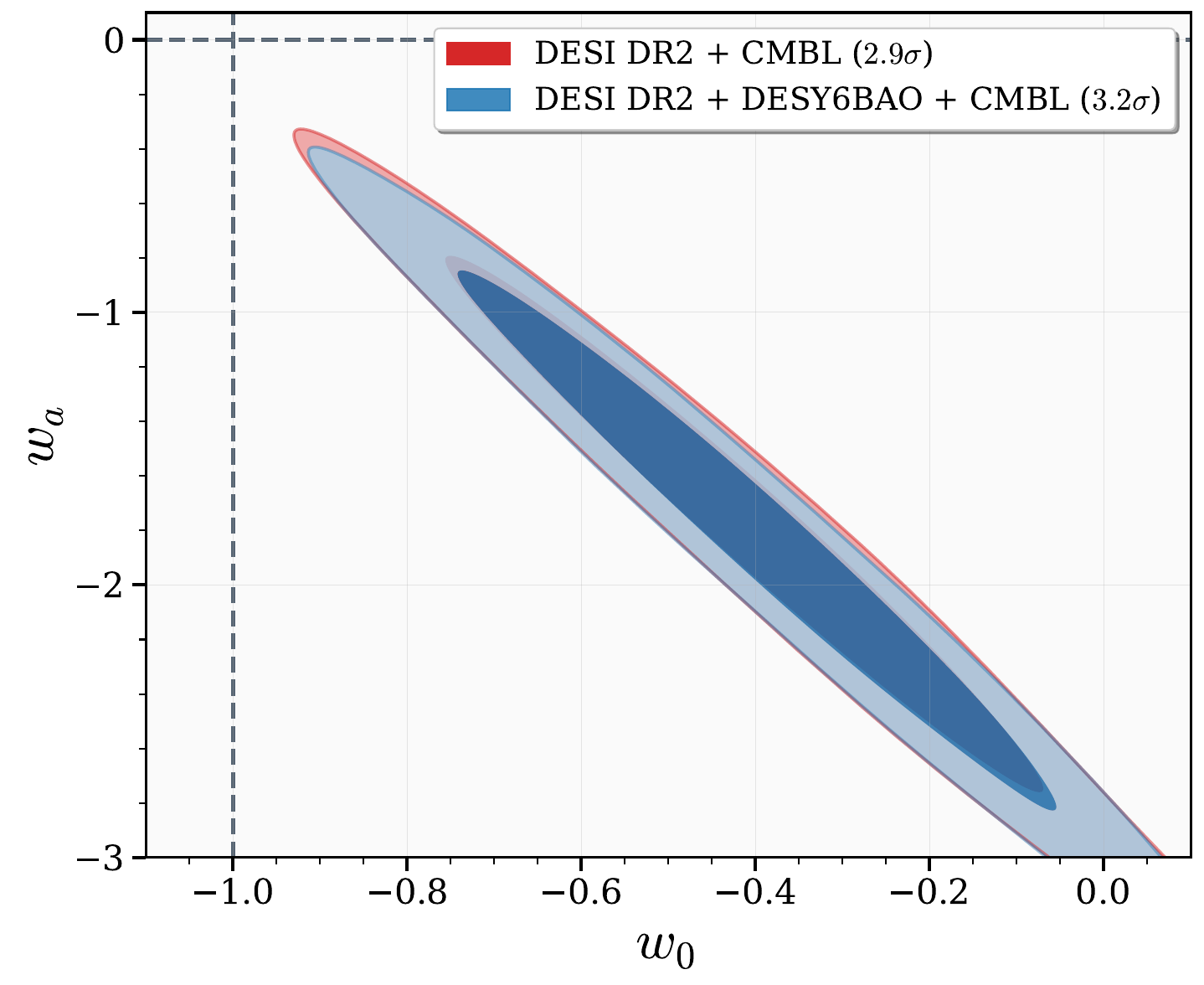} 
    \vspace{-6pt} 
    \caption{No-SN dataset combination plots: 68\% and 95\% marginalised posterior constraints in the $w_0$--$w_a$ plane for the flat \wowa\ model. Each of these combinations favors the quadrant $w_0>-1$, $w_a<0$, and exhibits a preference of time-evolving EOS dynamical dark energy over a cosmological constant. The corresponding constraints and significance details are provided in \cref{tab:no_SN}.}
    \label{fig:no_SN}
\end{figure}

After DESI’s first paper \cite{DESI:2024mwx} announcing tantalizing results that favor a dynamical dark energy with a time varying equation of state over a cosmological constant of a the \lcdm\ model, some criticism was raised about the fact that the level of significance when using combined constraints depends on the supernova dataset used in the combinations. In \cref{fig:no_SN} and \cref{tab:no_SN}, we show that the tension with the cosmological constant persists even in cases where we do not use any supernova data at all. Indeed without supernova data, the combinations DESI DR2 + CMBL and DESI DR2 + DESY6BAO + CMBL are in tension with a cosmological constant at the levels of 2.9$\sigma$ and 3.2$\sigma$, respectively. Certainly, the disparity between the supernova datasets and their systematics is an issue that will need more attention and a resolution within the corresponding scientific community, and some of that is in its way \cite{DES:2025tir}, however, the results here show that the dynamical dark energy preference over \lcdm\ stands even without any supernova data. 

\begin{table*} [h]
    \centering
    \resizebox{\linewidth}{!}{
   \begin{tabular}{lccccccc}
    \hline
    Model/Datasets & $\Omega_m$ & $H_0$ [km s$^{-1}$ Mpc$^{-1}$] & $w_0$ & $w_a$ & $\dchisq$ & n-$\sigma $   & $\Delta$(DIC) \\
\hline
\wowa  \\
DESI + CMBL & $0.352\pm 0.021$ & $63.7^{+1.7}_{-2.1}$ & $-0.42\pm 0.21$ & $-1.74\pm 0.59$ & -11.0 & 2.9\footnote{due to sampling-noise instability propagated to the \texttt{iminuit} minimizer in this combination, this number differs from the value in \cite{DESI:2025zgx}, but we note that the marginalized posterior constraints we obtain here using other systems and pipelines are practically identical to those of \cite{DESI:2025zgx}.
We also verified that starting the \texttt{iminuit} minimizer from the MAP points from the DESI publicly-released chains does provide the 3.1-$\sigma$ reported in \cite{DESI:2025zgx}.}

& -9.5\\
DESI + DESY6BAO + CMBL & $0.352\pm 0.021$ & $63.7^{+1.7}_{-2.1}$ & $-0.41^{+0.23}_{-0.20}$ & $-1.79\pm 0.59$ & -13.5 & 3.2 & -9.2 \\
\hline
\lcdm  \\
DESI + CMBL & $0.3026\pm 0.0036$ & $68.18\pm 0.28$ & - & - & - & - & - \\
DESI + DESY6BAO + CMBL & $0.3018\pm 0.0036$ & $68.24\pm 0.28$ & - & - & - & - & - \\
\hline
    \end{tabular}
    }
    \caption{No-SN dataset combination results. Marginalized posterior means for cosmological parameters quoted along with 68\% credible intervals or one-sided 68\% upper limits for the \wowa\ model, as well as for the corresponding \lcdm\ model.  We also list the $\dchisq$,  n-$\sigma $ significance and $\Delta$(DIC), as described in the text, to compare the \wowa\ over the corresponding \lcdm\ model. 
    \label{tab:no_SN}
    }
\end{table*}

\clearpage  
\subsection{\label{sec:no_CMB}NO CMB Data}
%
\begin{figure} 
    \centering
    \includegraphics[width = 0.50\columnwidth]{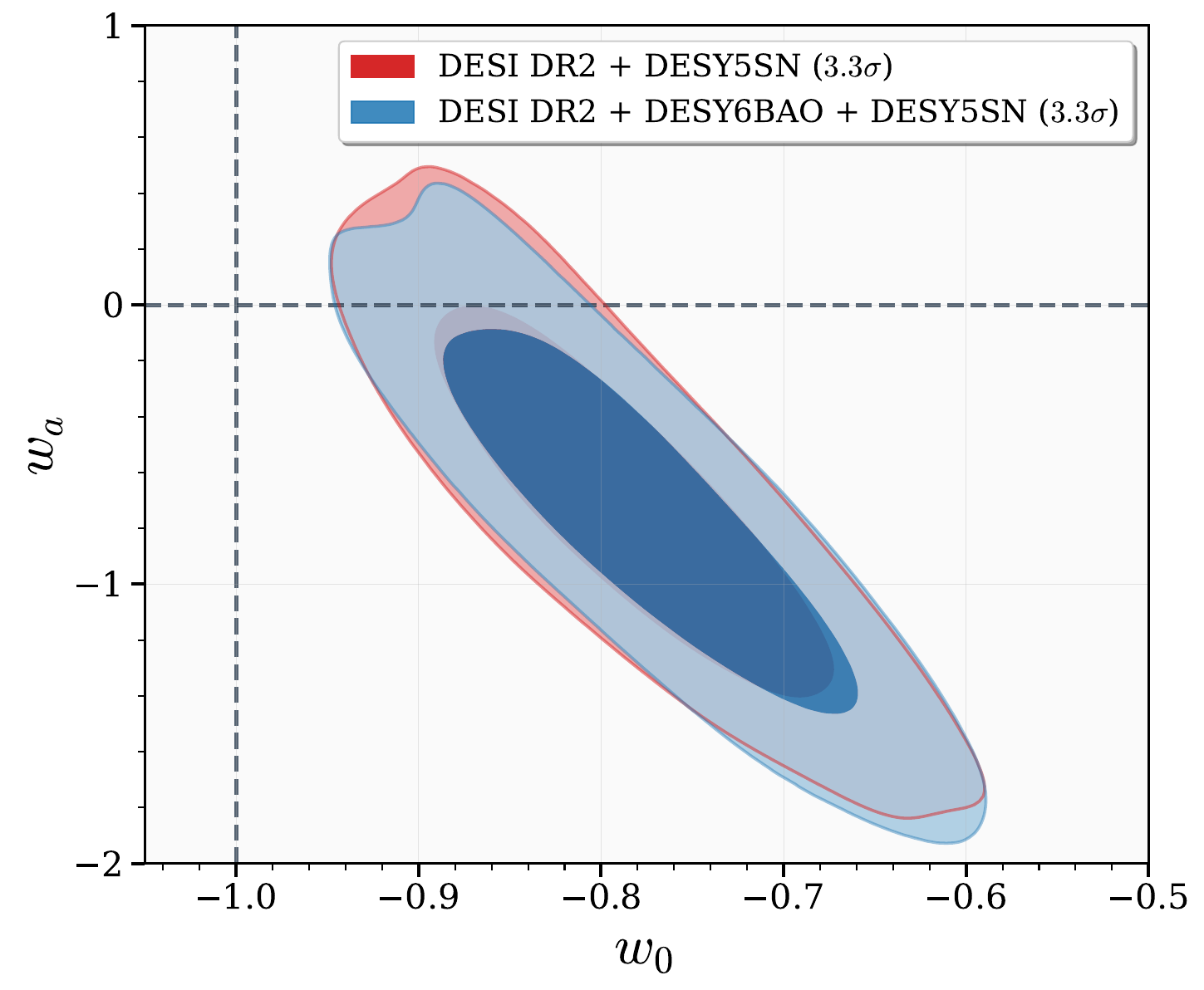} 
    \caption{No-CMB dataset combination plots: 68\% and 95\% marginalised posterior constraints in the $w_0$--$w_a$ plane for the flat \wowa\ model. Each of the combinations primarily favors the quadrant $w_0>-1$, $w_a<0$, and exhibits a preference of dynamical dark energy with a time-evolving equation of state over a cosmological constant. The corresponding constraints and significance details are provided in \cref{tab:no_CMB}.}
    \label{fig:no_CMB}
\end{figure}

 CMB does not significantly constrain the parameters of the equation of state of dark energy, however when added to other datasets, like supernova, it is able to break degeneracies between cosmological parameters and allows to tighten the constraint on the equation of state of dark energy. CMB also provides constraints from very high redshift and the case was discussed in previous papers \cite{DESI:2025zgx} that it is trying to fit simultaneously low, medium and the high redshift data that brings further tension within the \lcdm\ model, on for example the matter density parameter, but such tension is resolved within a dynamical dark energy model \wowa. We show here in \cref{fig:no_CMB} and \cref{tab:no_CMB} that a tension with the cosmological constant persists even in data combinations where no CMB data is used. Indeed, the combinations DESI DR2 + DESY5SN and DESI DR2 + DESY6BAO + DESY5 SN favor dynamical dark energy over a cosmological constant at the levels of 3.26$\sigma$ and 3.34$\sigma$, respectively, and regardless of any CMB data.  
\vspace{15pt}
\begin{table*} [h]
    \centering
    \resizebox{\linewidth}{!}{
   \begin{tabular}{lccccccc}
    \hline
    Model/Datasets & $\Omega_m$ &  $hr_\mathrm{d}$ [Mpc] & $w_0$ & $w_a$ & $\dchisq$ & n-$\sigma $   & $\Delta$(DIC) \\
\hline
\wowa  \\
DESI + DESY5SN & $0.318^{+0.017}_{-0.011}$ & $98.55\pm 0.85$ & $-0.784^{+0.065}_{-0.076}$ & $-0.70\pm 0.48$ & -13.6 & 3.3 & -9.6 \\
DESI + DESY6BAO + DESY5SN & $0.319^{+0.016}_{-0.012}$ & $98.60\pm 0.84$ & $-0.777^{+0.069}_{-0.077}$ & $-0.74\pm 0.47$ & -14.2 & 3.3 \footnote{because of our rounding to only a single decimal number, these two numbers appear to be the same but they actually correspond to 3.26$\sigma$ and 3.34$\sigma$, respectively.} & -10.4 \\
\hline
\lcdm  \\
DESI + DESY5SN & $0.3100\pm 0.0085$ & $100.56\pm 0.70$ & - & - & - & - & - \\
DESI + DESY6BAO + DESY5SN & $0.3093\pm 0.0082$ & $100.66\pm 0.68$ & - & - & - & - & - \\
\hline
    \end{tabular}
    }
    \caption{No-CMB dataset combination results. Marginalized posterior means for cosmological parameters quoted along with 68\% credible intervals or one-sided 68\% upper limits for the \wowa\ model, as well as for the corresponding \lcdm\ model.  We also list the $\dchisq$,  n-$\sigma $ significance and $\Delta$(DIC), as described in the text, to compare the \wowa\ over the corresponding \lcdm\ model. 
    \label{tab:no_CMB}
    }
\end{table*}
\clearpage 
\subsection{\label{sec:level2}NO BAO Data}

\begin{figure} 
    \centering
    \includegraphics[width = 0.50\columnwidth]{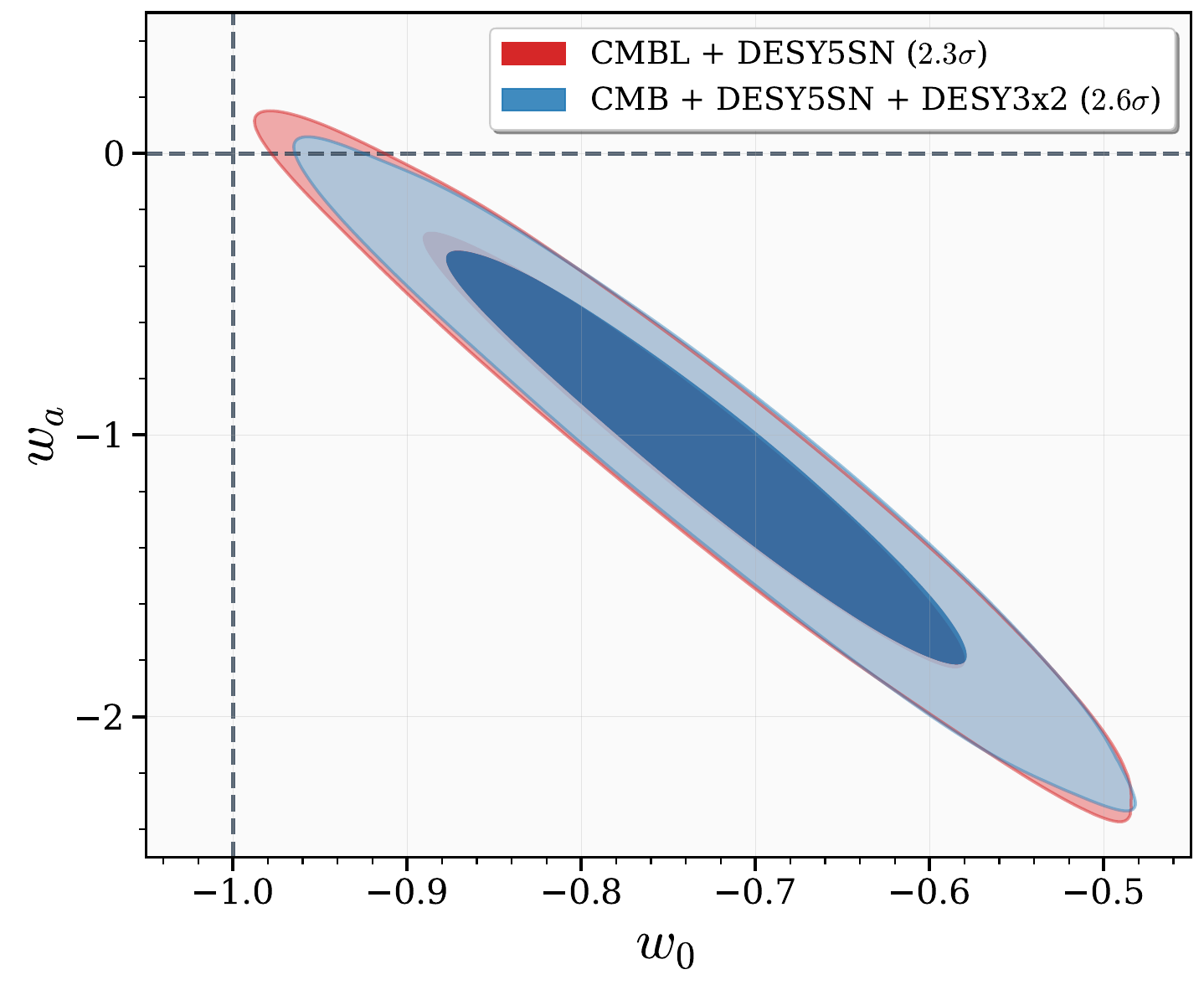} 
    \caption{No-BAO dataset combination plots: 68\% and 95\% marginalised posterior constraints in the $w_0$--$w_a$ plane for the flat \wowa\ model. Each of these combinations majoritarily favors the quadrant $w_0>-1$, $w_a<0$, and exhibits a preference of time-evolving EOS dynamical dark energy over a cosmological constant. The corresponding constraints and significance details are provided in \cref{tab:no_BAO}.}
    \label{fig:no_BAO}
\end{figure}

 DESI BAO data played a crucial role in providing significant results for dynamical dark energy with a time-evolving equation of state when combined with other datasets \cite{DESI:2024mwx,DESI:2025zpo, DESI:2025zgx}. DES BAO \cite{DESY6BAO} confirmed later the results from DESI DR1 and found consistent results for preference of \wowa. It is to be noted that such a trend for \wowa\ model was apparent in the completed SDSS BAO survey \cite{eBOSS:2020yzd} with a small significance.  However, it is interesting to see in \cref{fig:no_BAO} and \cref{tab:no_BAO} that a preference for such a dynamical dark energy over \lcdm\ stands for other data set combinations even without any BAO data at all. Indeed, as shown in \cref{fig:no_BAO} and \cref{tab:no_BAO}, the combinations CMBL + DESY5SN and  CMB + DESY5SN + DES3x2pts show a departure from a cosmological constant at the levels of 2.3$\sigma$ and 2.6$\sigma$, respectively, and that regardless of any BAO data.

\vspace{15pt}

\begin{table*} [h]
    \centering
    \resizebox{\linewidth}{!}{
   \begin{tabular}{lccccccc}
    \hline
    Model/Datasets & $\Omega_m$ & $H_0$ [km s$^{-1}$ Mpc$^{-1}$] & $w_0$ & $w_a$ & $\dchisq$ & n-$\sigma $   & $\Delta$(DIC) \\
\hline
\wowa \\
\hline
CMBL + DESY5SN & $0.3160^{+0.0095}_{-0.011}$ & $67.2\pm 1.0$ & $-0.73\pm 0.10$ & $-1.06\pm 0.51$ & -7.9 & 2.3 & -4.3 \\
CMB + DESY5SN + DESY3x2 & $0.3132^{+0.0082}_{-0.0092}$ & $67.37\pm 0.90$ & $-0.722\pm 0.098$ & $-1.09^{+0.51}_{-0.47}$ & -9.6 & 2.6 & -5.5 \\
\hline
\lcdm  \\
CMBL + DESY5SN & $0.3218\pm 0.0062$ & $66.79\pm 0.44$ & - & - & - & - & - \\
CMB + DESY5SN + DESY3x2 & $0.3152\pm 0.0054$ & $67.24\pm 0.39$ & - & - & - & - & - \\
\hline
    \end{tabular}
    }
    \caption{No-BAO dataset combination results. Marginalized posterior means for cosmological parameters quoted along with 68\% credible intervals or one-sided 68\% upper limits for the \wowa\ model, as well as for the corresponding \lcdm\ model.  We also list the $\dchisq$,  n-$\sigma $ significance and $\Delta$(DIC), as described in the text, to compare the \wowa\ over the corresponding \lcdm\ model. 
    \label{tab:no_BAO}
    }
\end{table*}

\clearpage  
\section{Putting it all together}\label{sec:combining}
\vspace{-10pt} 
\begin{figure} 
    \centering
    \includegraphics[width =\columnwidth]{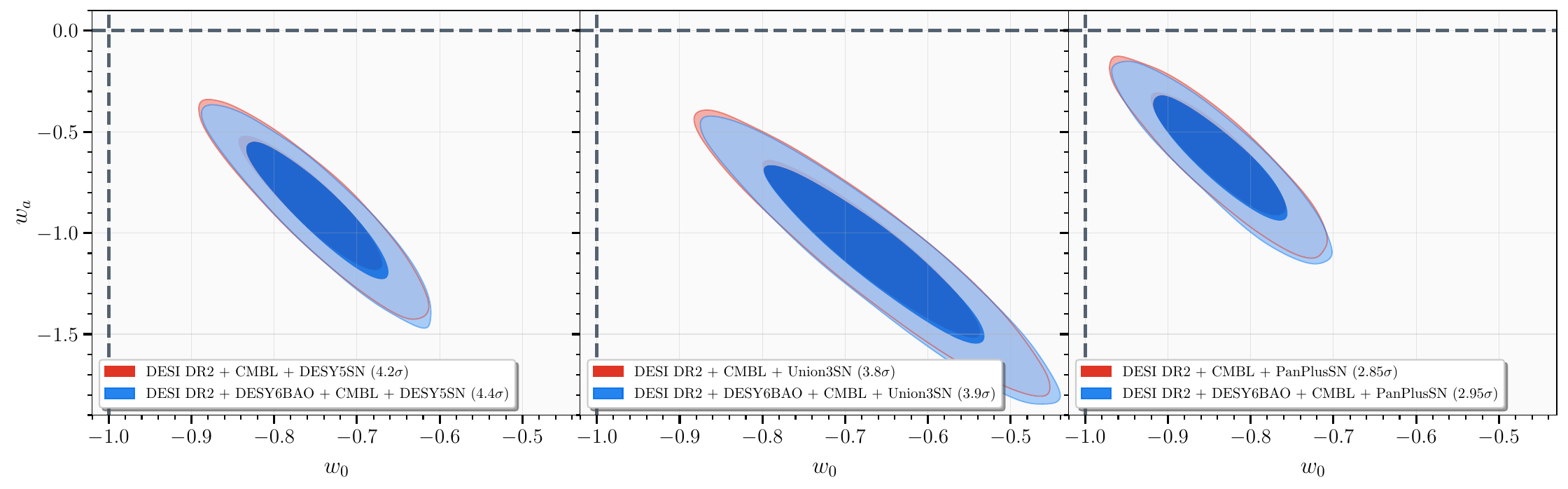} 
    \vspace{-8pt} 
    \caption{Larger combinations of dataset result plots: 68\% and 95\% marginalised posterior constraints in the $w_0$--$w_a$ plane for the flat \wowa\ model. Each of these combinations favors the quadrant $w_0>-1$, $w_a<0$, and exhibits a preference of time-evolving EOS dynamical dark energy over a cosmological constant. The corresponding constraints and significance details are provided in \cref{tab:HS_comb}.}
    \label{fig:HS_comb}
\end{figure}
 We show in this subsection results with the combinations including all three types of data sets (BAO, CMB and SN) that provide constraints on the equation of state of dark energy. As shown in \cref{fig:HS_comb} and \cref{tab:HS_comb}, each dataset combination provide contour plots and numbers that are clearly located in the quadrant $w_0>-1$, $w_a<0$ and in tension with the cosmological constant. In each panel, the addition of DES BAO to DESI DR2 BAO in the combinations moves the contour plot slightly further away from the cosmological constant and adding to the significance of the results.  We obtain constraints from the new combinations DESI + DESY6BAO + CMBL + DESY5SN, DESI + DESY6BAO + CMBL + Union3SN, and DESI + DESY6BAO + CMBL + PanPlusSN giving 4.4$\sigma$,  3.9$\sigma$, and 2.9$\sigma$ significance levels, respectively.  

\begin{figure}[h]
\centering
\begin{subfigure}
\centering
\includegraphics[width=\textwidth]{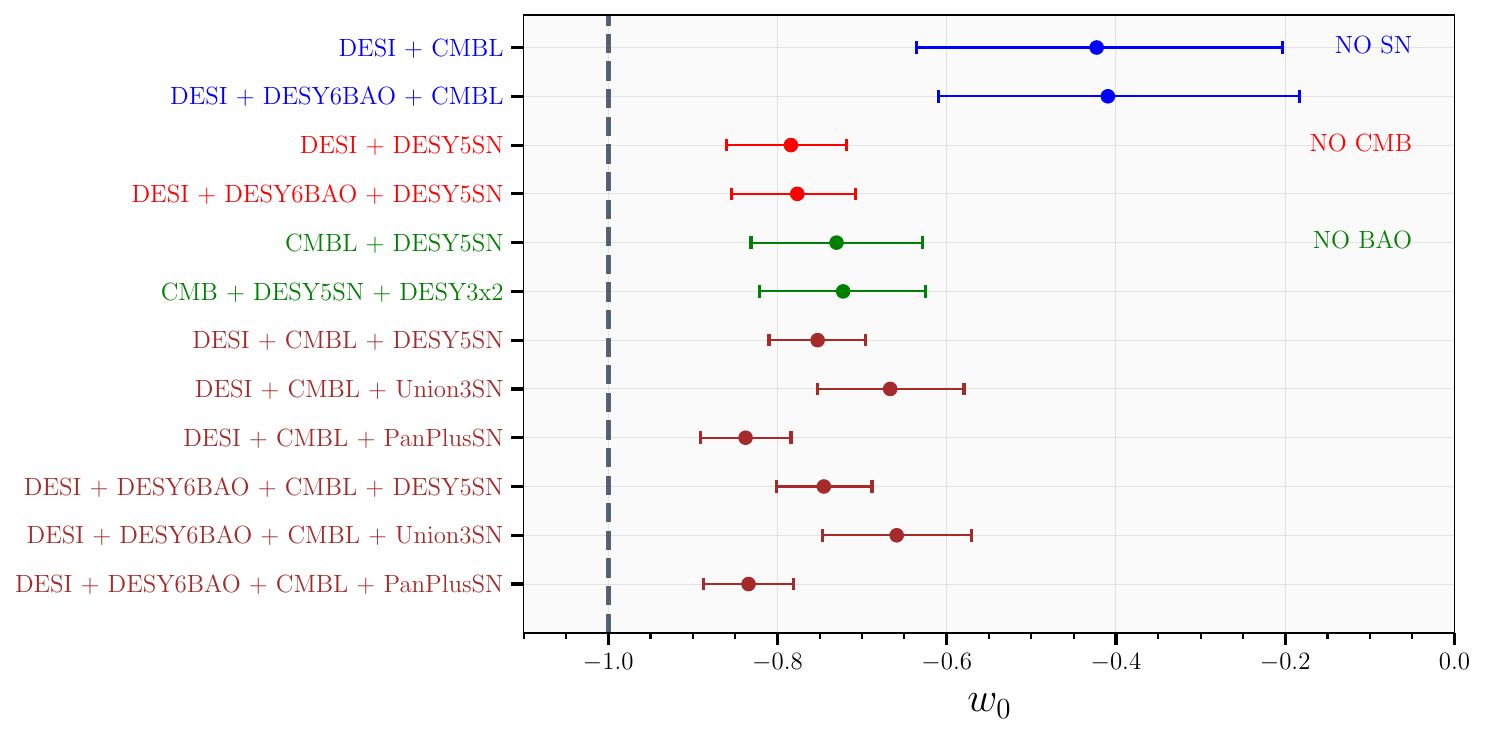}
\end{subfigure}
\vspace{0.5cm}
\begin{subfigure}
\centering
\includegraphics[width=\textwidth]{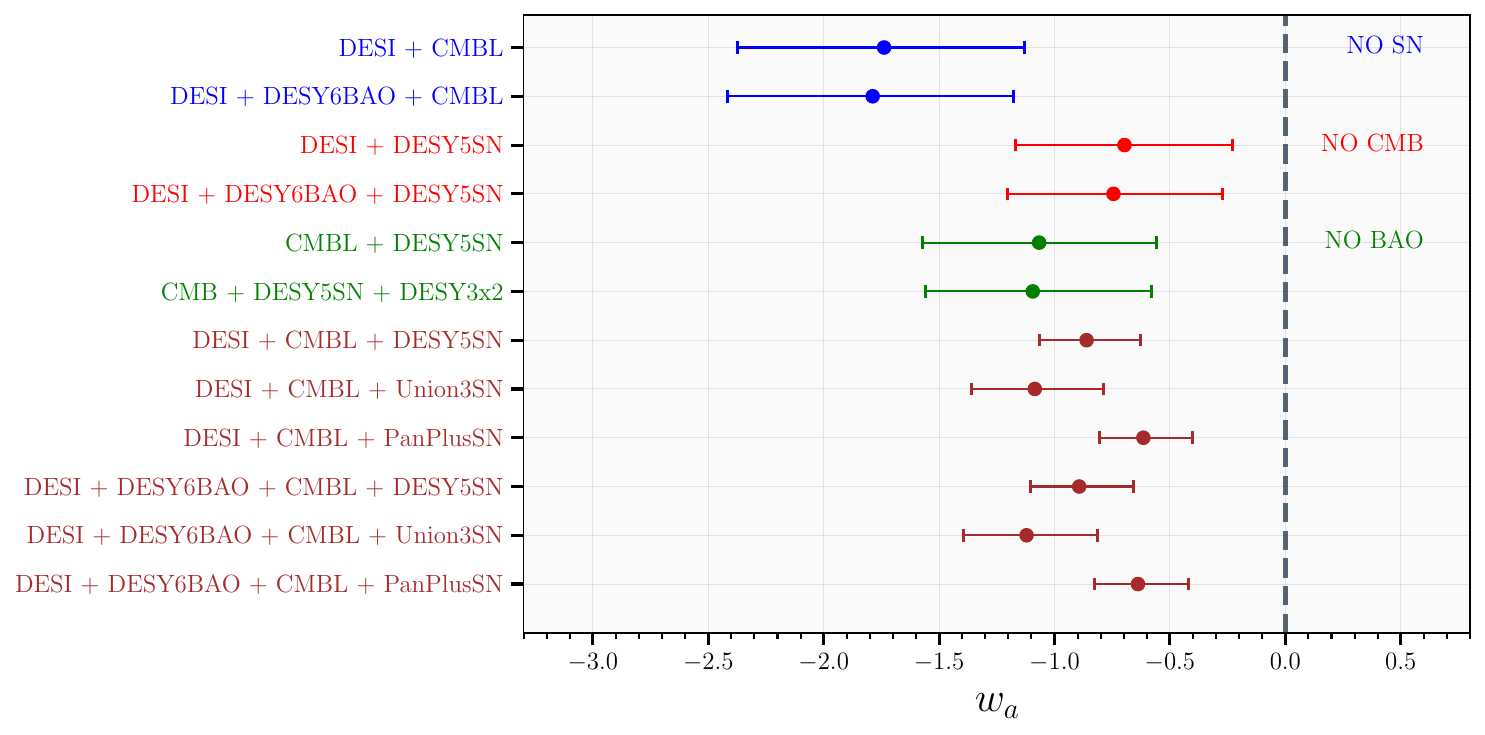}
\end{subfigure}
\caption{
Summary plot of the 1-D constraints on the equations of state parameters from the dataset combinations used providing the marginalized means and 68\% credible intervals on $w_0$ (top panel) and $w_a$ (bottom panel).  Data combinations organized in NO SN, NO CMB, NO BAO and then all 3 types combined. The vertical black dashed lines represent the $w_0=-1.0$ and $w_a=0.0$ of the cosmological constant. 
Each combination in the 6 pairs (or trios) at the top of each panel show a preference for the data combination of the \wowa\ model over the \lcdm\ model regardless of the type of data excluded. Below that, the combinations of the three types of data provide tighter constraints with higher significance for such a preference pattern as given in \cref{tab:HS_comb}. The two panels show an overall persistent portrait that is not in favor of the standard \lcdm\ model.  
}
\label{fig:wiskerplots_w0wa}
\end{figure}

\begin{figure} 
    \centering
    \includegraphics[width =\columnwidth]{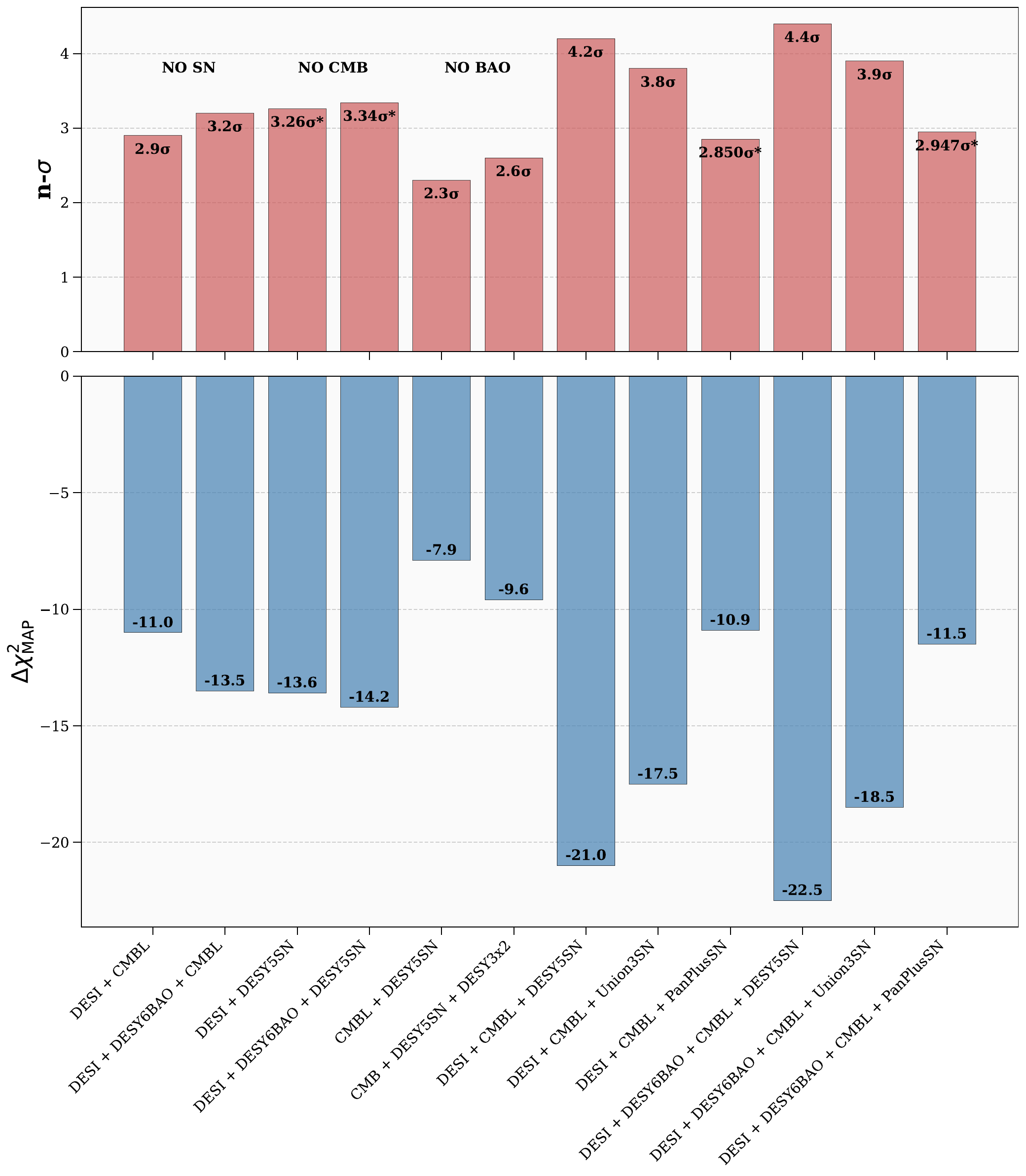} 
    \caption{n-$\sigma$ and $\Delta\chi^2_\mathrm{MAP}$ quantification of the level of preference of the \wowa\ model over the \lcdm\ model versus the data combination used. The calculation method of these quantities is described in section \cref{sec:backround}.  The datasets are organized as in the overall analysis with pairs (or trios) that have NO SN, NO CMB, or NO BAO, and then combinations including all three types of datasets. A persistent and consistent overall portrait is depicted showing preference of the data for the \wowa\ model over the \lcdm\ model regardless of any particular type of dataset.}
    \label{fig:significance}
\end{figure}

Furthermore, we illustrate our results from this section and the previous ones into a 1-D information Whisker plot showing marginalized means and 68\% credible intervals on $w_0$ and $w_a$ in \cref{fig:wiskerplots_w0wa} \footnote{Ref. \cite{Giare:2025pzu} provided a similar plot for dark energy EOS parameters with data available before DESI DR2},  as well as two bar chart plots showing the corresponding evidence levels in \cref{fig:significance}. In concert with the various 2-D contour plots we provided, the two panels in \cref{fig:wiskerplots_w0wa} and in \cref{fig:significance} draw an overall clear portrait showing that all the dataset combinations used here give constraints that have drifted away from a cosmological constant and are consistently in favor of a time-varying equation of state \wowa\ model over a \lcdm\ model.  

\begin{table*} [h]
    \centering
    \resizebox{\linewidth}{!}{
   \begin{tabular}{lccccccc}
    \hline
    Model/Datasets & $\Omega_m$ & $H_0$ [km s$^{-1}$ Mpc$^{-1}$] & $w_0$ & $w_a$ & $\dchisq$ & n-$\sigma $   & $\Delta$(DIC) \\
\hline
\wowa \\
\hline
DESI + CMBL + DESY5SN & $0.3190\pm 0.0057$ & $66.75\pm 0.57$ & $-0.753\pm 0.057$ & $-0.86^{+0.23}_{-0.21}$ & -21.0 & 4.2 & -17.5 \\
DESI + CMBL + Union3SN & $0.3275\pm 0.0086$ & $65.91\pm 0.84$ & $-0.667\pm 0.087$ & $-1.09\pm 0.29$ & -17.5 & 3.8 & -13.6 \\
DESI + CMBL + PanPlusSN & $0.3115\pm 0.0057$ & $67.50\pm 0.60$ & $-0.838\pm 0.054$ & $-0.62^{+0.21}_{-0.19}$ & -10.9 & 2.9 & -6.4 \\
DESI + DESY6BAO + CMBL + DESY5SN & $0.3185\pm 0.0056$ & $66.79\pm 0.56$ & $-0.745\pm 0.057$ & $-0.89^{+0.24}_{-0.21}$ & -22.5 & 4.4 & -18.3 \\
DESI + DESY6BAO + CMBL + Union3SN & $0.3270\pm 0.0086$ & $65.95\pm 0.84$ & $-0.659\pm 0.088$ & $-1.12^{+0.31}_{-0.27}$ & -18.5 & 3.9 & -14.2 \\
DESI + DESY6BAO + CMBL + PanPlusSN & $0.3105\pm 0.0056$ & $67.59\pm 0.59$ & $-0.834\pm 0.054$ & $-0.64^{+0.22}_{-0.19}$ & -11.5 & 2.9\footnote{because of our rounding to only a single decimal number, this numbers appears to be the same as its homologous above but they actually correspond to 2.850$\sigma$ and 2.947$\sigma$, respectively.} & 2.3 \\
\hline
\lcdm  \\
DESI + CMBL + DESY5SN & $0.3051\pm 0.0036$ & $67.99\pm 0.28$ & - & - & - & - & - \\
DESI + CMBL + Union3SN & $0.3037\pm 0.0036$ & $68.09\pm 0.28$ & - & - & - & - & - \\
DESI + CMBL + PanPlusSN & $0.3038\pm 0.0035$ & $68.08\pm 0.27$ & - & - & - & - & - \\
DESI + DESY6BAO + CMBL + DESY5SN & $0.3042\pm 0.0035$ & $68.06\pm 0.27$ & - & - & - & - & - \\
DESI + DESY6BAO + CMBL + Union3SN & $0.3028\pm 0.0035$ & $68.16\pm 0.28$ & - & - & - & - & - \\
DESI + DESY6BAO + CMBL + PanPlusSN & $0.3030\pm 0.0035$ & $68.15\pm 0.27$ & - & - & - & - & - \\
\hline
    \end{tabular}
    }
    \caption{
    Marginalized posterior means for cosmological parameters quoted along with 68\% credible intervals or one-sided 68\% upper limits for the \wowa\ model, as well as for the corresponding \lcdm\ model.  We also list the $\dchisq$,  n-$\sigma $ significance and $\Delta$(DIC), as described in the text, to compare the \wowa\ over the corresponding \lcdm\ model.  
    \label{tab:HS_comb}
    }
\end{table*}

\clearpage 

\clearpage

\clearpage
%
\section{\label{sec:conclusion}Conclusion}
%
In this paper, we derived new results and reproduced independently some existing ones about constraints on dark energy with a time-evolving equation of state using multiple combinations of datasets from  BAO, CMB (with and without lensing), supernovae, and cross/auto-correlations between galaxy positions and galaxy lensing (i.e. 3x2pts) from DES. 

Firstly, we organized and used pairs and trios of datasets into combinations while excluding one type of the datasets each time. We categorized such combinations as “NO SN”, “NO CMB”, and “NO BAO”. We find that, in each case, the data combinations show a preference for the \wowa\ with $w_0>-1$ and $w_a<0$ over the \lcdm\ model with a significance ranging from mild (2.3$\sigma$) to moderate (3.3$\sigma$) including new combinations such as DESI + DESY6BAO + CMBL and DESI + DESY6BAO + DESY5SN. The BAO addition from DES works in the same direction as that of DESI and increases the significance of the preference for the \wowa\ over \lcdm. It is worth noting that such a drift of the equation of state parameter constraints toward this preferred quadrant, $w_0>-1$ and $w_a<0$, was already noticeable in the SDSS complete survey analysis \cite{eBOSS:2020yzd} although with much less significance. 

While these levels of significance from pairs or trios are relatively smaller than the ones we provided in the second part that we recapitulate in the next paragraph, they do come from various combinations where a given dataset was excluded along with its possible systematic errors.  This first part serves the purpose of a validation of the reliability of the pattern and trend observed in these results regardless of a single type of dataset and its related systematics. 
Namely, a time-evolving dark energy preference over a cosmological constant stands even when supernovae datasets have been set aside altogether with, for instance, the combination DESI + DESY6BAO + CMB giving 3.2$\sigma$.
Similarly, the combination DESI + DESY6BAO + DESY5SN gives 3.3$\sigma$ without CMB, and CMB + DESY5SN + DES3x2pts gives 2.6$\sigma$ without any BAO. 
Moreover, it is worth mentioning that even without DESI at all, and without any supernovae datasets, the trend still persists and we find that DESY6BAO + CMBL yields 2.5$\sigma$ 
{(\cite{DES:2018ufa} found for this combination the evidence to be $>$ 3.0$\sigma$ using a different minimization methodology).}   
Therefore, this first part of the analysis shows that the persistent hints for a preference of a dynamical dark energy equation of state (either effective or true) over a cosmological constant cannot be attributed solely to a given dataset and this cannot be ignored on the premise of unidentified systematic effects in one given dataset. It would require an unlikely confluence of systematic effects in each of the independent datasets that would push results away from a cosmological constant and inexplicably all into the fourth quadrant of the equation of state where $w_0>-1$ and $w_a<0$. Although this is not completely impossible, it is highly improbable. 
This makes it hard to ignore or deny this pattern of preference for dynamical dark energy over a cosmological constant or to try to associate the result with a problem in a single dataset. 

Secondly, we used combinations that include all three types of datasets and derived results that show preference of the data for a time-evolving dark energy over \lcdm\ with larger statistical significance.  We verified that such datasets are mutually consistent within the \wowa\ models before combining them.  We find that the new combinations DESI + DESY6BAO + CMBL + DESY5SN, DESI + DESY6BAO + CMBL + Union3SN, DESI + DESY6BAO + CMBL + PanPlusSN give 4.4$\sigma$,  3.9$\sigma$, and 2.95$\sigma$ significance levels, respectively, improving on previous results. 

It is worth noting that it remains to be investigated what is the specific underlying nature of this dynamical dark component with such a time-varying effective (or true) equation of state, be it an unknown form of dark energy based on specific scalar fields, the result of interactions in the dark sector, a modification to general relativity at cosmic scales, or simply that the overall way we are putting the pieces together in the standard model of cosmology needs a major reconstruction. 

In sum, the combination of the first step supporting validation and reliability of these results since they persist regardless of any single type of dataset exclusion and their associated systematics, along with the second step showing high-significance results when such datasets are combined, builds a compelling overall case in support of a dynamical dark energy with a time-evolving equation of state over the cosmological constant of the \lcdm.  Although it is important to remain prudent while awaiting future data, the results here draw a clear portrait of a pattern and a trend that have become difficult to dismiss. The evidence and validation coming from various combinations of datasets raise the question of whether this is the beginning of the end for the reign of the \lcdm\ standard model of cosmology.  
\vspace{-19pt} 
\begin{acknowledgments}
\vspace{-11pt} 
MI acknowledges the use of the Texas Advanced Computing Center (TACC) at The University of Texas to perform all the MCMC runs, minimization runs, and calculations reported within this paper. We thank Cristhian Garcia Quintero for providing useful comments on the manuscript and Kristian Gonzalez for proof-reading it. 
MI acknowledges that this material is based upon work supported in part by the Department of Energy, Office of Science, under Award Number DE-SC0022184 and also in part by the U.S. National Science Foundation under grant AST2327245.
\end{acknowledgments}

\appendix

\section{Parameters and priors}

\begin{table}[h] 
    \centering
    \begin{tabular}{|llll|}
    \hline
    parameter & symbol & default & prior\\  
    \hline 
    \textbf{background-probes-only} &&&\\
    
    matter density& $\Omega_m$ &---& $\mathcal{U}[0.01, 0.99]$\\
    
    sound horizon times the normalized Hubble parameter & $\rd h$ \; ($\Mpc$) &---& $\mathcal{U}[10, 1000]$  \\
    
    Hubble parameter (after $\rd$ calibration) & $H_{0} \; (\kmsMpc)$ &---& $\mathcal{U}[20, 100]$  \\

     physical baryon density & $\ob$ &---& $\mathcal{U}[0.005, 0.1]$  \\
\hline 
    \textbf{CMB included} &&&\\
    
    physical cold dark matter density& $\ocdm$ &---& $\mathcal{U}[0.001, 0.99]$ \\
    
    physical baryon density & $\ob$ &---& $\mathcal{U}[0.005, 0.1]$ \\
    
     acoustic angular scale & $100 \theta_{\mathrm{MC}}$ &---& $\mathcal{U}[0.5, 10]$ \\
    
    amplitude of primordial scalar fluctuations& $\ln(10^{10} A_{s})$ &---& $\mathcal{U}[1.61, 3.91]$ \\
    
    spectral index & $n_{s}$ &---& $\mathcal{U}[0.8, 1.2]$ \\
    
    reionization optical depth & $\tau$ &---& $\mathcal{U}[0.01, 0.8]$ \\
    
    \hline 
    \textbf{time-evolving EOS dynamical dark energy extension} &&&\\
    present value of the equation of state of dark energy& $w_0$   & $-1$ & $\mathcal{U}[-3, 1]$ \\
    Minus slope of the EOS of dark energy& $w_{a}$ & $0$ & $\mathcal{U}[-3, 2]$ \\
    \hline
    \end{tabular}
    \caption{
    Cosmological parameters and corresponding default values and priors used in the analysis. Priors are flat in the ranges given. The parametrisation``background-probes-only" is employed when using BAO and SN data with NO CMB. ``CMB included" is where data from Planck and ACT are used in combination with other datasets. The same priors are used for those parameters when we use the time evolving EOS dark energy model extension. In addition to the flat priors on the EOS parameters $w_0$ and $w_a$ in the table, we also impose the requirement $w_0+w_a<0$ so that a period of high-redshift matter domination is enforced.
    }
    \label{tab:priors}
\end{table}

\bibliography{refs_key_paper,DESI2024,DDEComment, Leo,INSPIRE-CiteAll_filtered_sorted}

\end{document}